\documentclass[aps,prl,showpacs,twocolumn,groupedaddress]{revtex4-1}

\usepackage{amsmath}
\usepackage{cancel}
\usepackage{float}
\usepackage{ifsym}
\usepackage{color}
\usepackage{latexsym}
\usepackage{latexsym}
\usepackage{dcolumn}
\usepackage{epsfig}
\usepackage{amsfonts}
\usepackage{amsmath}
\usepackage{dsfont}
\usepackage{accents}
\usepackage[all,cmtip]{xy}

\usepackage{bm}
\usepackage{graphicx}
\usepackage{wasysym}
\usepackage{hyperref} 

\begin{document}

\title{Giant domain wall response of highly twinned ferroelastic materials}

\author{W. Schranz}
\email[]{wilfried.schranz@univie.ac.at}
\author{H. Kabelka}
\author{A. Sarras}
\author{M. Burock}
\affiliation{$^1$University of Vienna, Faculty of Physics, Boltzmanngasse 5, 1090 Wien, Austria}

\date{\today}

\begin{abstract}
Many ferroelastic crystals display at sufficiently low measurement frequencies a huge elastic softening below T$_c$ which is caused by domain wall motion. Materials range from perovskites to iron based superconductors and shape memory materials. We present a  model - based on Landau-Ginzburg theory including long range elastic interaction between needle shaped ferroelastic domains - to describe the observed superelastic softening. The theory predicts that the domain wall contribution to the elastic susceptibility is  different for improper and proper ferroelastic materials. A test of the theory against experimental data on SrTiO$_3$, KMnF$_3$, LaAlO$_3$,  La$_{1-x}$Nd$_x$P$_5$O$_{14}$ and NH$_4$HC$_2$O$_4$$\cdot\frac{1}{2}$H$_2$O yields excellent agreement.
\end{abstract}

\maketitle

The macroscopic response of materials depends often on structures spanning a broad range of length and time scales. Examples of such inhomogeneous structures are precursor clusters near structural phase transitions, domains and domain walls, interfaces at first order displacive or reconstructive phase transitions, etc. They have been reported in a wide range of functional materials. 
Prominent cases are the enhancement of piezoelectric response \cite{Hlinka} of nanotwinned BaTiO$_3$ for domain thickness below 50 nm, or the giant piezoelectric effect in ferroelectric relaxors \cite{Kutnjak} due to the high mobility of polar nanoregions \cite{Cohen}. Generally, domains and domain boundaries in ferroic crastals can act as structural elements for creating novel functional devices \cite{Salje2010}. The usefulness of domain boundaries in Domain Boundary Engineering depends on the time scale of their response. Static or pinned domain boundaries can be used as functional units since they can host functional properties (ferroelectric, superconducting, etc.) which are absent in the bulk \cite{Salje2009}. Mobile domain boundaries can lead to giant macroscopic responses, depending on the frequency of the changing external field.

In the present work we will focus on ferroelastic materials, which depending on the type of coupling between the primary order parameter $\eta$ and the strains $\varepsilon$ can be classified as proper, pseudo proper, improper or co-elastic ones \cite{Salje1993}. Very often ferroelastic crystals consist of a large number of domains which are separated by domain boundaries \cite{Tagantsev2011}. Due to mechanical compatibility these domain boundaries should be planar with well defined orientation. However, in real crystals very often needle or dagger shaped domains appear \cite{Salje1993}. There are still many open questions, e.g. concerning the stability of ferroelastic domains, their motion under an applied dynamic stress, the observed domain freezing \cite{Schranz2009} at sufficiently low temperature, etc.

Understanding the macroscopic behaviour of multidomain crystals is important for technological applications as well as for the interpretation of seismic signals of our Earth, since domain wall motion influences the low frequency elastic and anelastic behaviour of minerals at seismic frequencies (1-20 Hz). In recent years we have performed quite detailled low frequency (0.1-100 Hz) measurements in a number of perovskites \cite{Kityk2000,Schranz2009,Zhang2010} with improper ferroelastic phase transitions. In all these materials we found a huge elastic softening in the low symmetric ferroelastic phase due to the influence of domain wall motion. Calculation of the domain wall response to an applied dynamic stress is hampered, since ferroelastic domains are in general metastable or even unstable objects, in contrast to ferroelectric or ferromagnetic ones, where the competition between the domain wall energy and the deplarization or demagnetization field leads to a stable domain pattern. For this reason research of ferroelastic materials has been mainly focused on first order phase front \cite{Dec1993,Roytburd1998} - or substrate \cite{Alpay1998} - stabilized arrays of ferroelastic domains. However, it was shown previously \cite{Torres1982}, that at the end of ferroelastic needle shaped domains long range elastic stress fields appear - which is reminiscent of the stray fields in ferroelectric or ferromagnetic crystals. These stress fields stabilize an array of ferroelastic domains. Very recently  we have set up a Landau-Ginzburg free energy including long range elastic interactions between needle shaped domains to calculate the equilibrium domain width and the resulting macroscopic elastic response of a ferroelastic multidomain crystal \cite{Schranz2011,Schranz2012}, yielding perfect agreement with the data on improper ferroelastic perovskites.
In the present work we show, that this theory predicts the domain wall response to be very different for \textit{improper} as compared to \textit{proper} or \textit{pseudo proper} ferroelastic phase transitions.  We will demonstrate this on the examples of SrTiO$_3$\cite{Schranz2011} and KMnF$_3$ \cite{Schranz2009} (improper), La$_{1-x}$Nd$_x$P$_5$O$_{14}$ (proper)\cite{Huang1995} and NH$_4$HC$_2$O$_4$$\cdot\frac{1}{2}$H$_2$O (AHO, pseudo proper).
AHO single crystals were grown in our lab following the procedure of Godet, et al.\cite{Godet1987}. For the elastic measurements a dynamical mechanical analyzer (DMA7, Perkin Elmer) was used in three point bending geometry \cite{Schranz1997}. A dynamic force F$_{dyn}$ at f=0.1-50 Hz is superimposed to a static force F$_{stat}$, yielding both, real and imaginary parts of the complex elastic compliance S$^*$=S$'$+iS$''$ by measuring the corresponding dynamic amplitude u and phase angle $\delta$ , i.e.

\begin{equation}\label{eq:ThreePointBending}
\frac{u}{F} = \frac{S(\mathbf{q})}{4t}\left(\frac{L}{h}\right)^{3}\left(1+\frac{3h^{2}}{2L^{2}}\cdot\frac{S(\mathbf{p,q})}{S(\mathbf{q})}\right)e^{i\delta}
\end{equation}

where L, h and t are the length, span and thickness of the sample bar, respectively. $\mathbf{q}$ points along the long axes of the sample bar and $\mathbf{p}$ is the direction of the applied force F. $S(\mathbf{q})$ and $S(\mathbf{p,q})$ are the longitudinal and shear components of the elastic compliance tensor, respectively. The corresponding Young's modulus is defined as Y:=1/S. Since in most cases $\frac{h}{L}\approx10^{-2}$, the second part in Eq.(\ref{eq:ThreePointBending}) can be usually  neglected. However, approaching a proper ferroelastic transition, $S(\mathbf{p,q})\rightarrow\infty$ and the second term in Eq.(\ref{eq:ThreePointBending}) can easily overcome the first one, making three point bending an ideal method to measure ferroelastic softening, i.e. $Y(\mathbf{p,q})\rightarrow0$ for $T\rightarrow T_{c}$.

Supposing a multidomain ferroelastic crystal is exposed to a stress $\sigma=\sigma_{stat}+\delta\sigma_{dyn}$ the width  $x_{+}$ of energetically preferred domains increases, whereas the width $x_{-}$ of oppressed domains shrinks. This motion of domain walls leads to a macroscopic deformation $\varepsilon^{DW}$ given as \cite{Schranz2011}

\begin{equation}\label{eq:EpsDomainwall}
\varepsilon^{DW}=\varepsilon_{s}\left(1-\frac{x_{-}}{2d}\right)
\end{equation}

where $\varepsilon_{s}$ is the spontaneous strain and $d$ the average width of domains. Using Eq.(\ref{eq:EpsDomainwall}), one obtains for the domain wall contribution to the elastic compliance

\begin{equation}\label{eq:DWcompliance}
\Delta S^{DW}=\frac{\partial\varepsilon^{DW}}{\partial\sigma}=-\frac{\varepsilon_s}{2d}\left(\frac{\partial x_-}{\partial\sigma}\right)
\end{equation}

Taking into account the long range elastic interactions between needle shaped domains and adding the repulsion between domain walls of finite thickness $w$ to the Landau free energy density $f_L$

\begin{equation}\label{eq:Landaufreeenergy}
f_{L}=A(T-T_c)\eta^{2}+B\eta^{4}+C\eta^{6}
\end{equation}

one obtains the free energy density in the presence of applied stress as \cite{Schranz2011}

\begin{widetext}
\begin{equation}\label{eq:Freenergy}
f(\sigma)=f_L + k\varepsilon_{s}^{2}\left(x_{+}+x_{-}\right) +
\frac{1}{x_{+}+x_{-}}\left[E_{w}+b\cdot exp(-x_{+}/w)+ b\cdot exp(-x_{-}/w)\right] +
2\varepsilon\sigma\frac{x_{+}}{x_{+}+x_{-}}
\end{equation}
\end{widetext}

where $E_w>0$ is the domain wall energy and $b>0$ describes the repulsion of domain walls of finite width $w$ and A, B, C are temperature independent Landau coefficients.  The equilibrium conditions $\frac{\partial f(\sigma)}{\partial x_{\pm}}=0$  lead to two transcendental equations which can be used \cite{Schranz2011} to calculate $x_{+}(\sigma,T)$ and $x_{-}(\sigma,T)$ and finally also $\Delta S^{DW}(\sigma,T)$. Here we focus on the elastic anomalies at very small applied stresses, i.e. for $\sigma\rightarrow 0$, yielding $x_{+}(0,T)=x_{-}(0,T)=d(T)$ and

\begin{equation}\label{eq:finalDeltaSDW}
\Delta S^{DW}=\frac{\varepsilon_{s}^{2}w^{2}}{2bd}\cdot exp(2d/w)
\end{equation}

Eq.(\ref{eq:finalDeltaSDW}) is an extension of the classical results \cite{Sidorkin1998}, where zero thickness of domain walls was assumed leading to a domain wall contribution to the corresponding elastic compliance $\Delta S^{DW}\propto \varepsilon_{s}^{2}/d$. The exponential term in Eq.(\ref{eq:finalDeltaSDW}) does not significantly contribute to the elastic anomaly since it is weakly temperature dependent, approaching unity at $T_c$. Thus we will omit this term for further considerations.
Most importantly, in the present model $\Delta S^{DW}(T)$ is modified by the domain wall width squared $w^2(T)$. The temperature dependence of the domain wall width was measured \cite{Chrosch1999} e.g. for LaAlO$_3$ yielding a temperature dependence of $w\propto (T_c-T)^{-1}$. For a Landau-Ginzburg Free energy expansion up to sixth order and a one component order parameter $\eta$ it was shown \cite{Lajzerowicz1981} that

\begin{equation}\label{eq:DWwidth}
w^{2}=\frac{2g}{B\eta^{2}+2C\eta^{3}}
\end{equation}

where  $g$ is the coefficient of the order parameter gradient term. The temperature dependence of the spontaneous strain in Eq.(\ref{eq:finalDeltaSDW}) depends crucially on the nature of the phase transition, i.e. on the coupling between the order parameter $\eta$ and the strain variables $\varepsilon$, i.e. $\varepsilon_{s}\propto \eta^{2}(T)$  for \textit{improper} and $\varepsilon_{s}\propto \eta(T)$ for \textit{pseudo proper} ferroelastic phase transitions \cite{Salje1993}. For \textit{proper} ferroelastic phase transitions the order parameter is identical with the spontaneous strain, i.e. $\varepsilon_{s}\propto(T_c-T)^{1/2}$  for a 2-4 Landau free energy expansion.

Now we can use Eq.(\ref{eq:finalDeltaSDW}) and Eq.(\ref{eq:DWwidth}) to calculate the domain wall contributions to the elastic compliance. For \textit{improper} ferroelastic materials one obtains:

\begin{equation}\label{eq:DWcontimproper}
\Delta S^{DW} \propto \frac{N_w \eta^{2}}{B+2C\eta}
\end{equation}

where $N_w\propto 1/d$ is the number of domain walls in the sample.

For \textit{proper} ferroeleastics we can write:

\begin{equation}\label{eq:DWcontproper}
\Delta S^{DW} \propto \frac{N_w}{B+2C\varepsilon_{s}}
\end{equation}

and for the \textit{pseudo proper} ones we get:

\begin{equation}\label{eq:DWcontpseudo proper}
\Delta S^{DW} \propto \frac{N_w}{B+2C\eta}
\end{equation}

\begin{figure}
\includegraphics[scale=1.3, angle=90, clip=True]{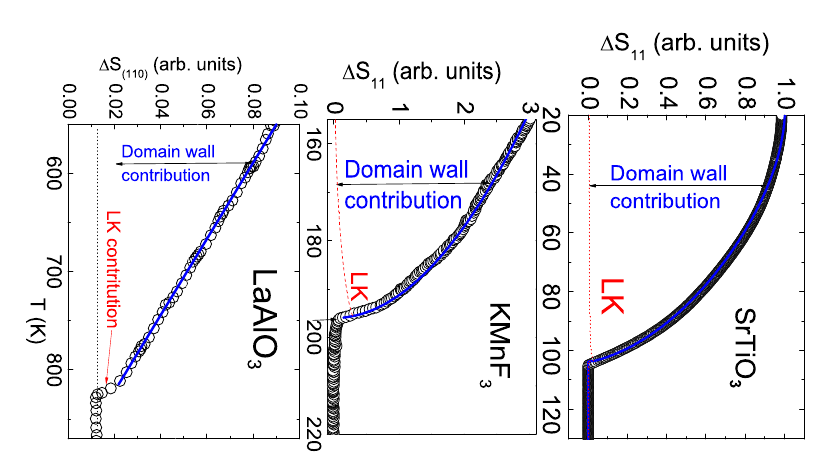}
\caption{DMA data of the anomalous parts of the elastic compliances of SrTiO$_3$, KMnF$_3$ and LaAlO$_3$ (points) and corresponding fits (blue lines) using Eq.(\ref{eq:DWcontimproper}). LK is the elastic anomaly calculated for a monodomain crystal, taking into account the order parameter strain coupling in the corresponding Landau free energy.}
\label{fig1}
\end{figure}

Equations (\ref{eq:DWcontimproper})-(\ref{eq:DWcontpseudo proper}) imply that the domain wall contributions to the elastic compliance are very different for the various cases. For \textit{improper} ferroelastics the anomaly in $\Delta S^{DW}$ is mainly determined by the temperature dependence of the number of domain walls $N_w(T)$ and the order parameter squared $\eta^2(T)$, whereas for \textit{pseudo proper} and \textit{proper} ferroelastics it is dominated by $N_w(T)$ only. In the following we will demonstrate, that these equations describe the experimental data of many rather different ferroelastic materials very well. Fig.\ref{fig1} shows the temperature dependencies of the compliances of three perovskites measured by DMA \cite{Kityk2000,Schranz2009,Harrison2002}. Both, SrTiO$_3$ and KMnF$_3$ undergo a cubic-tetragonal phase transition of second \cite{Salje1998} and weakly first order \cite{Hayward2000}, respectively, described by a Landau free energy expansion up to sixth order (2-4-6 Landau potential), i.e. $B,C\neq 0$ in Eq.(\ref{eq:Landaufreeenergy}). LaAlO$_3$ undergoes a second order phase transition from rhombohedral to monoclinic structure, described by 2-4 Landau potential, i.e. with $C=0$.
In all these cases the behaviour of $\Delta S^{DW}$ is governed by the temperature dependence of the order parameter only, since the number of domains was found to be constant in the whole measured temperature range. Note that the temperature dependencies of the order parameters in SrTiO$_3$ \cite{Salje1998}, KMnF$_3$ \cite{Hayward2000} and LaAlO$_3$ \cite{Harrison2002} were determined very detailed, so there is not much freedom for fitting these data. Nevertheless, as Fig.\ref{fig1} shows Eq.(\ref{eq:DWcontimproper}) yields excellent fits for all three cases. A very similar behaviour has also been found for Pb$_3$(PO$_4$)$_2$ \cite{Harrison2004}. It exhibits a first order rhombohedral-monoclinic phase transition, described by a 2-4-6 Landau potential, leading to a nonlinear Domain wall contribution to the elastic compliance similar as for the other 2-4-6 perovskites. The most important result for this part is, that for all these \textit{improper} ferroelastic cases $\Delta S^{DW} \rightarrow 0$ as $T \rightarrow T_c$.

\begin{figure}
\includegraphics[scale=0.8, angle=0, clip=True]{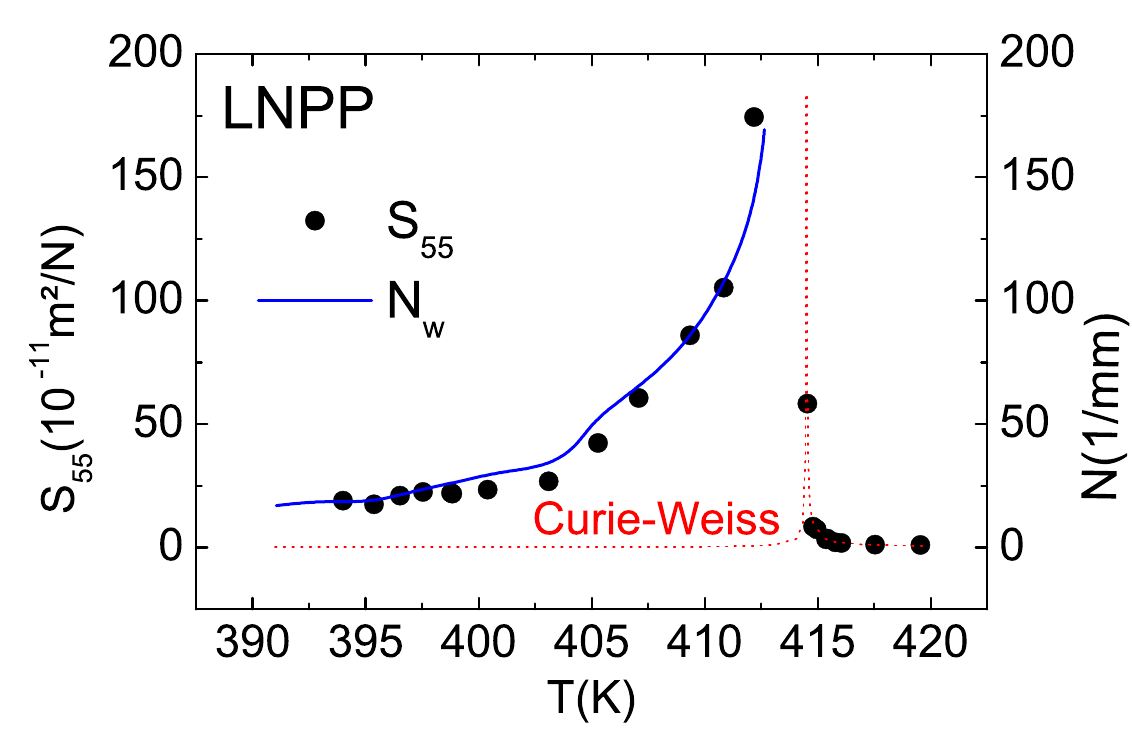}
\caption{Temperature dependence of the shear elastic compliance S$_{55}$ (black points) of LNPP (from Ref.\cite{Wang1987}) fitted with Eq.(\ref{eq:DWcontproper}) using the experimental data of N$_w$(T) (blue line). The red dotted line displays the Curie-Weiss law expected for a single domain crystal.}
\label{fig2}
\end{figure}

Now we will focus on the pseudo proper or proper ferroelastic cases. La$_{1-x}$Nd$_x$P$_5$O$_{14}$ (LNPP) \cite{Wang1987} as well as NdP$_5$O$_{14}$ (NPP) \cite{Huang1995} undergo a second order proper ferroelastic phase transition at T$_c \approx$ 414 K from orthorhombic $Pncm$ to monoclinic $P2_1/c$. A fourth order expansion of the Landau potential in terms of the shear strain $\varepsilon_5$ as primary order parameter describes the phase transition very well \cite{Wang1987,Huang1995}. Below T$_c$ the crystal splits into ferroelastic domains forming a regular stripe pattern with opposite shear strain $\pm \varepsilon_5$. The number of domains increases from about 20/mm to about 180/mm  for LNPP and even more drastically from about 100/mm to 1500/mm for NPP when approaching T$_c$ from below. The transverse elastic constant C$_{55}$ has been measured in LNPP by a resonator method at several tens kHz \cite{Wang1987}. Fig.\ref{fig2} displays the shear compliance $S_{55}=C^{-1}_{55}$, which should exhibit a Curie-Wiss temperature dependence (red dotted line), i.e. $S_{55} \propto |T_c-T|^{-1}$ in the single domain state. In contrast the experimental data (filled points) show significant domain wall contributions below T$_c$, which excellently scale with the measured number of domain walls, as predicted from Eq.(\ref{eq:DWcontproper}).

Finally we show the results for NH$_4$HC$_2$O$_4$$\cdot\frac{1}{2}$H$_2$O. Ammonium hydrogen oxalate hemihydrate (AHO) exhibits a pseudo proper ferroelastic phase transition of second order at T$_c$=146 K \cite{Keller1982} from $Pmnb$ to $P2_1/n$. The bilinear coupling between the primary order parameter $\eta$ and the shear strain $\varepsilon_{5}$ leads to a softening of the shear elastic constant $C_{55}$ which indeed was previously measured by Brillouin scattering \cite{Benoit1986}.
The corresponding inverse shear elastic constant $S_{55}=1/C_{55}$ of a single domain crystal is expected to follow a Curie-Weiss type law \cite{Benoit1986} given as

\begin{eqnarray}\label{eq:Curie-Weiss-type}
S_{55}^{+} &=& S^{0}_{55} + \frac{T_c-T_0}{T-T_c}   \qquad T>T_c \nonumber\\
S_{55}^{-} &=& S^{0}_{55} + \frac{T_c-T_0}{2(T_c-T)}   \qquad T<T_c
\end{eqnarray}

where T$_c -$T$_0>0$, due to the bilinear coupling between $\eta$ and $\varepsilon_5$.

\begin{figure}
\includegraphics[scale=0.8, angle=0, clip=True]{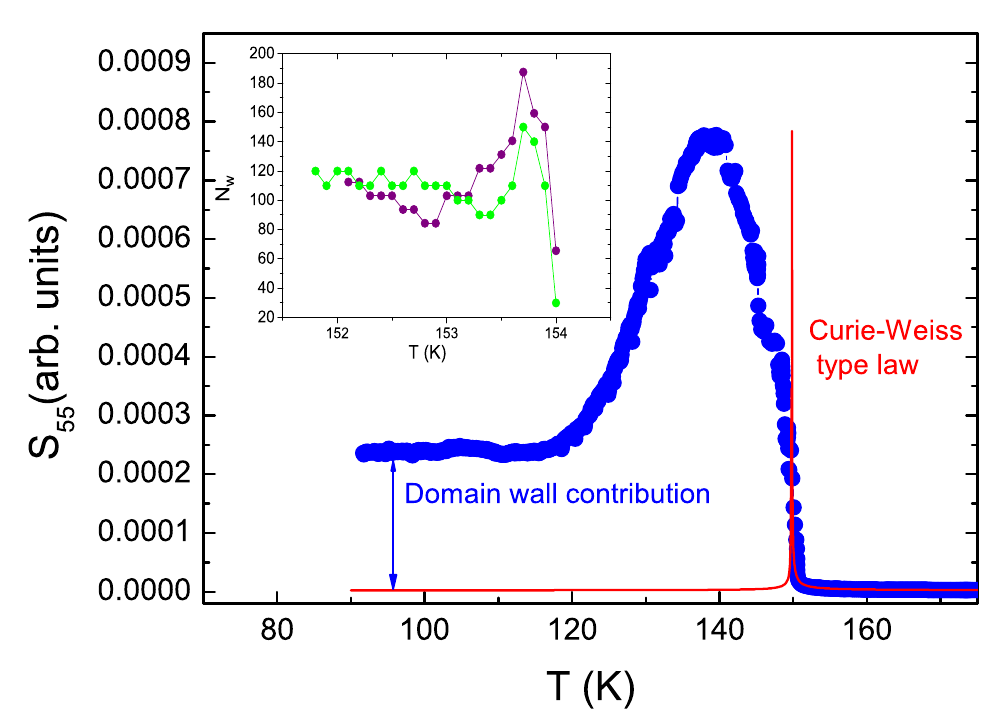}
    \caption{Temperature dependence of the shear elastic compliance S$_{55}$ of AHO measured with DMA7 in Three Point Bending geometry (blue points). The red line shows the Curie-Weiss type law Eq.(\ref{eq:Curie-Weiss-type}) of a nomodomain crystal. The inset shows the temperature dependence of the number of domain walls $N_w$ as determined from optical polarizing microscope measurements of two different AHO samples.}
   \label{fig3}
\end{figure}

Fig.\ref{fig3} shows the temperature dependence of S$_{55}$ of AHO measured by DMA in Three Point Bending Geometry at f=1Hz, which is in sharp contrast with the expected single domain behaviour.

 The measured DMA data are described by Eq.(\ref{eq:Curie-Weiss-type}) only \textit{above} T$_c$. In the ferroelastic phase the measured data exceed the Curie-Weiss type anomaly by several orders of magnitude similar as for LNPP (Fig.\ref{fig2}). The difference is due to domain wall motion.
 According to Eq.(\ref{eq:DWcontpseudo proper}) the anomaly of S$_{55}^{DW}$ is dominated by the temperature dependence of $N_w$ which increases with $T\rightarrow T_c$ (see insert of Fig.\ref{fig3}) in excellent agreement with the measured compliance data.

In conclusion we have shown, that for calculation of the macroscopic elastic response of multidomain crystals the finite width of domain walls and its temperature dependence has to be taken into account. In doing so, we have obtained excellent agreement with experimental data for quite a number of \textit{improper}, \textit{proper} and  \textit{pseudo proper} ferroelastic materials. It is expected, that these effects of domain wall thickness may be of similar importance also for other ferroic and multiferroic materials including also shape memory materials \cite{Ren2010}.

Financial support by the Austrian Science Fund (FWF) P23982-N20 is gratefully acknowledged.


\begin{thebibliography}{12}
\bibitem{Hlinka} J. Hlinka, P. Ondrejkovic and P. Marton, Nanotechnology \textbf{20}, 105709 (2009).
\bibitem{Kutnjak}Z. Kutnjak, R. Blinc and J. Petzelt, Nature (London) \textbf{441}, 956 (2006).
\bibitem{Cohen} R.E. Cohen, Nature (London) \textbf{441}, 941 (2006).
\bibitem{Salje2010} E.K.H. Salje, ChemPhysChem \textbf{11}, 940 (2010).
\bibitem{Salje2009} E. Salje and H. Zhang, Phase Transitions \textbf{82}, 452 (2009).
\bibitem{Salje1993} E.K.H. Salje, \textit{Phase Transitions in Ferroelastic and Coelastic Crystals} (Cambridge
  University Press, 1993).
\bibitem{Tagantsev2011} A.K. Tagantsev, L.E. Cross and J. Fousek, \textit{Domains in Ferroic Crystals and Thin Films} (Springer, 2010).
\bibitem{Schranz2009} W. Schranz, P. Sondergeld, A.V. Kityk and E.K.H. Salje, Phys. Rev. B \textbf{80}, 094110 (2009).
\bibitem{Kityk2000} A.V. Kityk, W. Schranz, P. Sondergeld, D. Havlik, E.K.H. Salje and J.F. Scott, Phys. Rev.
  B \textbf{61}, 946 (2000).
\bibitem{Zhang2010} Z. Zhang, J. Koppensteiner, W. Schranz, J.B. Betts, A. Migliori and M.A. Carpenter, Phys. Rev. B \textbf{82}, 014113(2010).
\bibitem{Dec1993} J. Dec, Phase Transitions \textbf{45}, 35 (1993).
\bibitem{Roytburd1998} A.L. Roytburd, J. Appl. Phys. \textbf{83}, 228 (1998); 83, 239 (1998).
\bibitem{Alpay1998} S. Pamir Alpay and A.L. Roytburd, J. Appl. Phys. \textbf{83}, 4714 (1998).
\bibitem{Torres1982}J. Torrés, C. Roucau and R. Ayroles, phys. stat. sol. (a) \textbf{70}, 193 (1982).
\bibitem{Schranz2011}W. Schranz, Phys. Rev. B \textbf{83}, 094120 (2011).
\bibitem{Schranz2012} W. Schranz, H. Kabelka and A. Tröster, Ferroelectrics \textbf{426}, 242 (2012).
\bibitem{Huang1995} X.R. Huang, S.S. Jiang, X.B. Hu, X.Y. Wu, W. Zeng, D. Feng and J.Y. Wang,
    Phys. Rev. B \textbf{52}, 9932 (1995).
\bibitem{Godet1987} J.L. Godet, M. Krauzman, J.P. Mathieu, H. Poulet and N. Toupry, Journal de
     Physique 48   no. 5, \textbf{809} (1987).
\bibitem{Schranz1997} W. Schranz, Phase Transitions \textbf{64}, 103 (1997).
\bibitem{Sonin1970} A.S. Sonin and B.A. Strukov, \textit{Introduction to Ferroelectricity} [in Russian] (Moscow 1970).
\bibitem{Sidorkin1998} A.S. Sidorkin, J. Appl. Phys. \textbf{83}, 3762 (1998).
\bibitem{Chrosch1999} J. Chrosch and E.K.H. Salje, J. Appl. Phys. \textbf{85}, 722 (1999).
\bibitem{Lajzerowicz1981} J. Lajzerowicz, Ferroelectrics \textbf{35}, 219 (1981).
\bibitem{Harrison2002} R. Harrison and S.A.T. Redfern, Phys. Earth and Planet. Int. \textbf{134}, 253 (2002).
\bibitem{Salje1998} E.K.H. Salje, M.C. Gallardo, J. Jim\'{e}nez, F.J. Romero and J. del Cerro, J. Phys.: Condens. Matter \textbf{10}, 5535 (1998).
\bibitem{Hayward2000} S.A. Hayward, F.J. Romero, M.C. Gallardo, J. del Cerro, A. Gibaud and E.K.H. Salje, J. Phys.: Condens. Matter \textbf{12}, 1133 (2000).
\bibitem{Harrison2004} R.J. Harrison, S.A.T. Redfern and U. Bismayer, Mineral. Mag. \textbf{68}, 839 (2004).
\bibitem{Wang1987} Y. Wang, W. Sun, X. Chen, H. Shen and B. Lu, phys. stat. sol. (a) \textbf{102}, 279 (1987).
\bibitem{Keller1982} H.J. Keller, D. Kucharczyk and H. K\"{u}ppers, Z. Krist. \textbf{158}, 221 (1982).
\bibitem{Benoit1986} J.P. Benoit, J. Berger, M Krauzman and J.L. Godet, J. Physique \textbf{47}, 815 (1986).
\bibitem{Ren2010} X. Ren, Y. Wang, Y. Zhou, Z. Zhang, D. Wang, G. Fan, K. Otsuka, T. Suzuki, Y. Ji,
J. Zhang, Y. Tian, S. Hou and X. Ding, Phil. Mag. \textbf{90}, 141 (2010).


\end{thebibliography}
\end{document}